\documentclass[twocolumn,twoside,slac_two]{revtex4}
\usepackage{graphicx}
\usepackage{fancyhdr}
\newcommand{\pt}{\ensuremath{p_T}}
\def\etmis{\ensuremath{\not \hspace{-0.3em} E_T}} 
\newcommand{\gev}  {\ensuremath{\mathrm{ GeV}}}
\newcommand{\gevc}{\ensuremath{\mathrm{ GeV/c}}}
\newcommand{\gevcs}{\ensuremath{\mathrm{ GeV/c^2}}}
\newcommand{\invpb}{\ensuremath{\mathrm{pb^{-1}}}}
\newcommand{\invfb}{\ensuremath{\mathrm{fb^{-1}}}}

\begin{document}

\title{Top physics at the Tevatron Run II}

%

\author{Boris Tuchming}
\affiliation{CEA Saclay, Dapnia/Spp, 91191 Gif-Sur-Yvette, France}

\begin{abstract}

The $p\bar p$ collider Tevatron, with its high center of mass energy of 2 TeV
is presently the unique top quark factory in the world.
Thousands of top-quarks have been produced during Run IIa.
This gives the opportunity to study the production and the properties of the heaviest known fundamental particle. This article will summarize a sample of recent top quark physics results obtained at the Tevatron.


\end{abstract}

\maketitle

\thispagestyle{fancy}


\section{Introduction}


The top quark  is the heaviest known fundamental fermion of the Standard Model (SM) with a mass of~$\simeq 175$~\gevcs, close to the  electroweak breaking scale. 
This unexpected large mass delayed for several years its discovery.
It occurred  only 11 years ago at the $p \bar p$
collider Tevatron~\cite{topdiscovery-D0,topdiscovery-CDF}, during the so called Run I~(1992-1996).

The high mass of the top quark implies a large coupling to the Higgs boson ($Y_{top}\simeq 1$).
This makes think it could play a particular role in the electroweak breaking mechanism and may be the gatekeeper to physics beyond the SM.
It also causes large effect to the $W$ boson propagation through virtual quantum effects,
so that precision measurements of the top mass together with the $W$ mass put strong constraints to the yet to be discovered Higgs boson.
Besides, this quark  has a very short lifetime and decays before it hadronizes. 
Therefore the bare quark properties are transfered to his decay products and are not hidden in the spectroscopy that we observe for the other
heavy flavor quarks.

Top physics can be studied at several levels.
First the production mechanism can be examined, for example by looking at the production kinematics or searching for a resonant production.
Secondly the particle properties, can be studied, such as its mass, width, lifetime, spin and charge.
Finally the decay properties and the couplings can be studied by looking at the decay products, their kinematics or  searching for rare decays.

As the top discovery is recent and only a few events were recorded during Run I,
top physics still need to be studied in details.
The only existing top factory at this time,
is the Tevatron, running at  1.96 TeV  since the beginning of Run II in spring 2001.
One of the main physics goal of the Tevatron experiments, CDF and D\O ,
is the study of the top  physics.

As of summer 2006, both experiments  have about 1 fb$^{-1}$ of data available
for physics analyzes, which is almost 8 times more than what was collected
during Run~I. Analysis presented in the following are based on smaller subsets, from
300 to 750~\invpb.
This is still a small fraction of what is expected to be recorded
by the end of Run II, in 2009: between 4 and 8 \invfb~\cite{bib-runii_lumi}.

\section{Production of top quarks at the Tevatron}
\subsection{Pair production}
\begin{figure}[hbt]
\centering
\includegraphics[width=30mm]{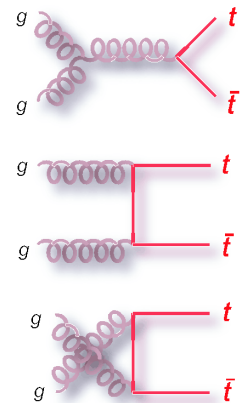}
\caption{Main diagrams for  top pair production.} \label{fig:diag-toppair}
\end{figure}

At the Tevatron, the top quarks are mainly produced by pair.
The  quark anti-quark annihilation contributes at the level of $\simeq 85\%$
while the gluon-gluon annihilation is only $\simeq 15\%$.
The main diagrams for these processes are displayed in Figure~\ref{fig:diag-toppair}.
The cross-section is predicted  to be  $6.7$~pb~\cite{Bonciani:1998vc,Cacciari:2003fi,Kidonakis:2003qe},
the level of uncertainty being $\simeq 12\%$.

Typically for $\simeq 1$~\invfb\ of data and for  a channel with a branching ratio of 30\% and an acceptance of 15\%,  300 events per experiment are expected to be observed.

\subsection{Electroweak single top production}
\label{sec:singletop}
\begin{figure}[hbt]
\centering
\includegraphics[width=35mm]{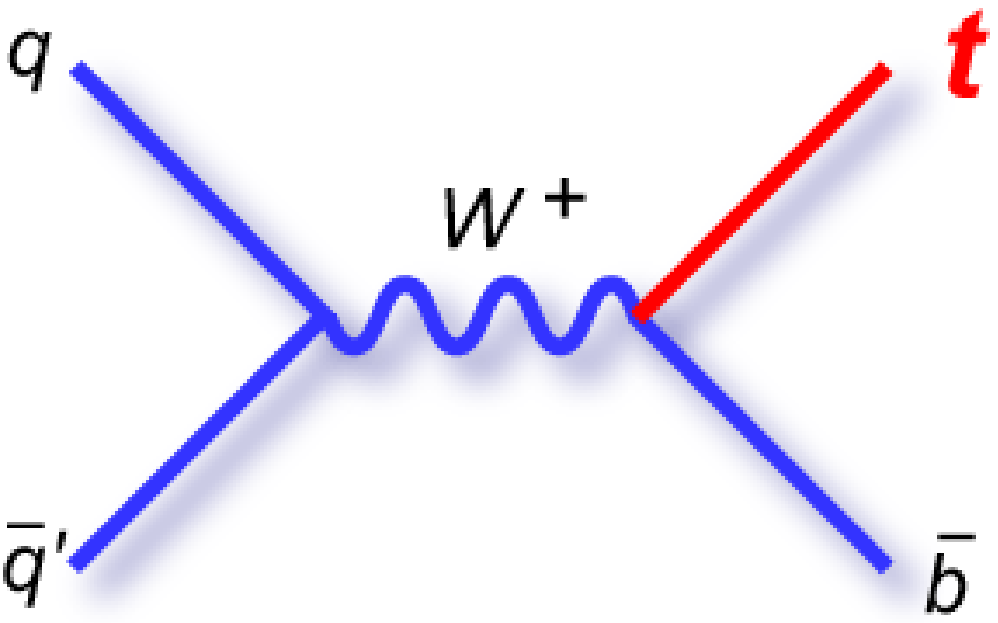}\hfill
\includegraphics[width=30mm]{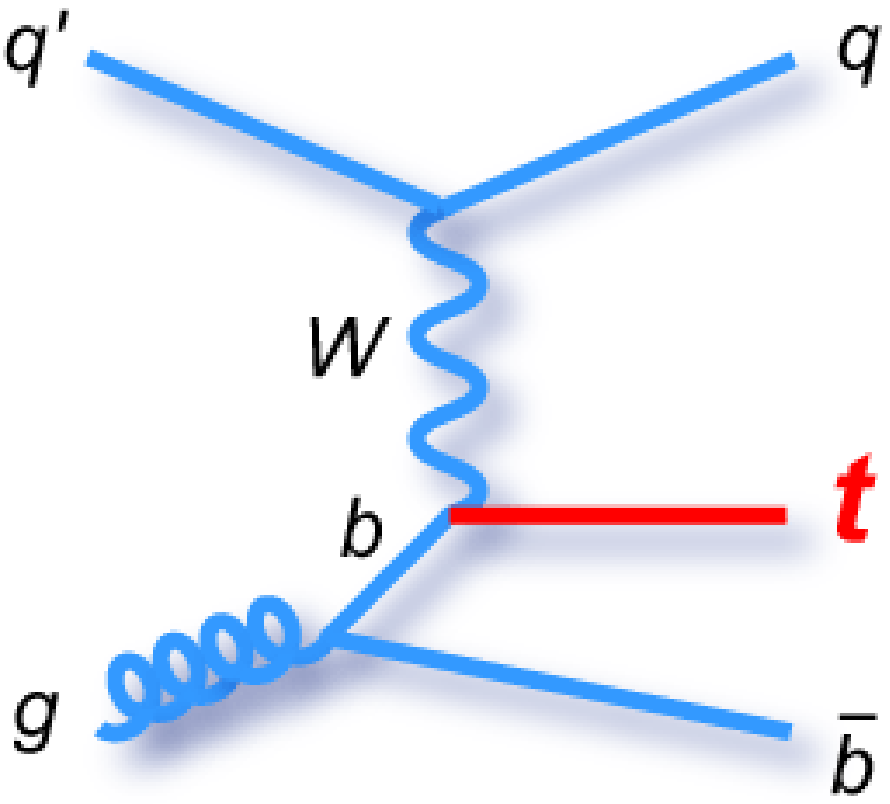}
\caption{Diagrams for single top production. The $s-$ and $t-$channels  have  different final states.} \label{fig:diag-singletop}
\end{figure}

Single top quarks can be produced via the electroweak coupling to the $W$ boson.
The main diagrams for these processes are displayed in Figure~\ref{fig:diag-singletop}.
Since they  have  a different final state, the  $s-$channel and $t-$channel are usually distinguished.
The $s-$ and the $t-$ channels 
have respectively cross sections of the order of
0.88~pb and 1.98~pb which is quite low.
Given that the background is quite high (there is only one top as a signature), this process has not been observed yet and ongoing analysis are searching for it.

\section{Top pair production}
\subsection{Signatures}
Within the Standard Model, the top decays into a  $W$ boson and a $b$ quark
almost 100\% of the time.
The different channels arise from the possible decays of the pair of $W$ bosons:
either $W\rightarrow e\nu$ ($\simeq 10.7\%$), or
$W\rightarrow \mu\nu$ ($\simeq 10.7\%$), or
$W\rightarrow \tau\nu$ ($\simeq 10.7\%$), or
$W\rightarrow  q\bar q'$ ($\simeq 69.0\%$).

The main channels are then:
\begin{itemize}
\item ``lepton + jets'' channels ($\simeq 30\%$)
correspond to events for which one $W$ decays hadronically and the other one decays into electrons or muons.  These channels have a moderate yield and a moderate background arising from  $W$+jets production, $Z$+jets production, or QCD processes. 

\item ``di-lepton'' channels ($\simeq 4.5\%$)
correspond to events for which the two $W$ decay  into electrons or muons.  These channels are very pure but have a small yield. The background is due to  di-bons events (mainly $WW$+jets production), $Z$+jets and also QCD processes. 

\item
``all jets channels'' ($\simeq 44\%$)
correspond to events where both $W$ bosons decay hadronically.
The yield is high, but the background arising from QCD multi-jet production is very large.

\item  
``tau channels'' ($\simeq 22\%$) arise from events where at least one of the $W$ decays into $\tau$. As the $\tau$ decays are hard to identify, especially in a hadronic environment as at Tevatron, these channels are weakly exploited.

\end{itemize}

Because of the high mass of the top quark,
the decay products have high momenta and large angular separations.
Reconstructing and identifying the production of top quarks demands
reconstruction and identification  of high transverse momenta ($\pt$) electrons, muons, jets, and the measurement of the missing transverse energy (\etmis). 
Good momentum  resolution  for these objects is  required and
the jet energy scale (JES) has also to be known with a good precision.

Identifying the $b$-jets is  an effective way of improving the purity of the selections.
The  $b$-tagging  makes use of the presence of secondary vertices and tracks with high impact parameters  involved by the decays of $b$-hadrons.
Typically for $\pt=50\ \gev$, a 50\%  efficiency per $b$-jet is achieved while the  mistag rate is about $0.5-1\%$.

Both CDF and D\O\   analyze their data using various methods, in the different channels. These numerous analysis are updated as the amount of data increases. Only a few examples and the typical selections are discussed in the following.

\subsection{Lepton+jets channels}
The signature consists of a central spherical energetic events, with a  high \pt\ lepton and a high \etmis. Four or more jets can be reconstructed. Among them two jets are expected to arise from $b$-quark.

The cuts to preselect the events are typically:
high \pt\ lepton trigger (lepton+jets trigger at D\O),
isolated lepton with $\pt>20$~\gevc, $\etmis>20$~GeV, 
4 jets with  $\pt>15$~\gevc.
Usually the scalar transverse energy $H_T$ defined as the sum
of transverse momenta of the reconstructed objects of the events,
$H_T=\sum_i
|\pt\,_i|$ can be used to improve purity, as $t\bar t$ events
are expected to have a higher $H_T$ than the background.

To discriminate between signal and background after these preselections, two main approaches are employed:
either the tagging of $b$-jets or topological cuts which maximize the use of the events kinematics.  
Each of these approaches is also a cross-check of the other one.

The acceptance is typically at the level of 10 to 20\% (for $N_{jets}\ge3$),
it is determined with the simulation (MC), corrected for DATA/MC weights
measured on independent samples.

The backgrounds arise mainly from $W+n$ jets, $Wjjj$, $Wbbj$, $Wccj$ (here $j$ stands for light partons, as opposed to heavy flavor, $b$ or $c$).
These backgrounds are determined with the MC, but as the yield suffers from
high theoretical uncertainties, these backgrounds are normalized to the data.
Non-$W$ background is mainly QCD where one jet fakes a lepton, it is determined on  the data.
The $WW$+jets background is determined using the MC.

\subsubsection{Lepton+jets with topological analysis at CDF}

\begin{figure}[thb]
\centering
\includegraphics[width=75mm]{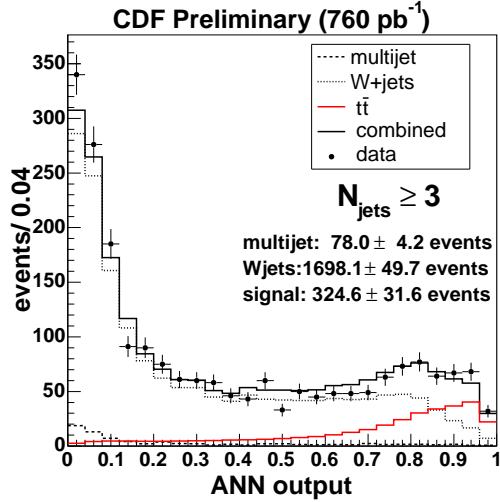}
\caption{distribution of the $ANN$ output of the selected events for the topological lepton+jets analysis at CDF.} \label{fig:cdf_ljet_nn}
\end{figure}
This analysis is based on an artificial neural network (ANN) which combines
seven kinematic variables to discriminate against the main background, $W$+jets.
A likelihood fit of the ANN output distribution allows to determine
the $t\bar t$  content of the selected events as quoted in Table~\ref{tab:cdf_ljet_nn_result}.
The ANN output distribution for selected events is shown in Figure~\ref{fig:cdf_ljet_nn}.
The measurement based on  760~\invpb\ of data reads:
$$\sigma(t\bar t)= 6.0\pm 0.6 \pm 0.9\ \mathrm{pb}.$$

\begin{table}
\begin{tabular}{|l|l|l|c|}
\hline
Sample        & Events       &  Fitted $t \bar t$     &$\sigma(t\bar t)$ \\
\hline
$W+\ge 3$-jets&  2102        &  $324.6 \pm 31.6$      & $6.0\pm 0.6 \pm 0.9$~pb\\
$W+\ge 4$-jets&   461        &  $166.0 \pm 22.1$      & $5.8\pm 0.8 \pm 1.3$~pb
\\
\hline
\end{tabular}
\caption{Results of the CDF topological letpon+jets analysis.}
\label{tab:cdf_ljet_nn_result}
\end{table}


The main systematic uncertainties for this measurement arise from the JES (8.3\%),
the $W$+jets modeling (10.2\%) and the luminosity measurement (5.8\%).

\subsubsection{Lepton+jets with $b$-tagging at CDF}

\begin{figure}[thb]
\centering
\includegraphics[width=75mm]{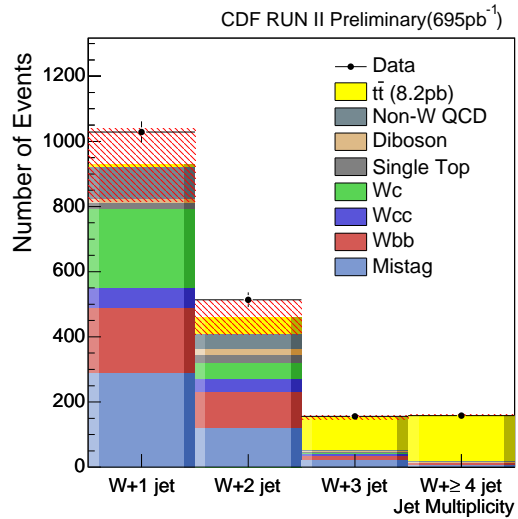}
\caption{Distribution of the jet multiplicity in lepton+jets data at CDF after requiring 1 b-tag and $H_T>200$~\gev.} \label{fig:cdf_ljets-btag}
\end{figure}

A secondary vertex tag is employ to identify  $b$-jets in this analysis based
on 695~pb$^{-1}$.
Events can have either $\ge 1$ b-tag or $\ge 2$ b-tag.
The jet multiplicity distribution for one b-tag events is shown in Figure~\ref{fig:cdf_ljets-btag}.
The cross-sections are measured from  the $3^\mathrm{rd}$ and $4^\mathrm{th}$ jet bin in $\ge 1$ b-tag events after the cut   $H_T>200$~\gev: $$
  \sigma(t\bar t)= 8.2\pm 0.6 (stat)\pm 1.0 (syst)\ \mathrm{pb}.$$

The main systematic uncertainties for this measurement arise from the b-tagging uncertainty (6.5\%)  the luminosity measurement (6.0\%) and the parton density function uncertainty (5.8\%).

The $t\bar t$ yield can be  also extracted from the jet multiplicity of $\ge 2$ b-tag events, but given the large  uncertainty this result is only a cross-check for the previous measurement:$$  
  \sigma(t\bar t)= 8.8 ^{+1.2}_{-1.1} (stat) ^{+2.0}_{-1.3} (syst)\ \mathrm{pb}.$$

\subsection{Di-leptons channels}
The  signature consists of a central spherical energetic events, with two  high \pt\ lepton and high \etmis. Two ore more jets can be reconstructed. Among them two jets are expected to arise from $b$-quark.

The cuts to preselect the events are typically:
high \pt\ lepton trigger (or di-lepton),
presence of isolated leptons with $\pt>20$~\gevc, $\etmis>20$~GeV, 
2 jets with  $\pt>15$~\gevc\ and
$H_T> 120-200$~GeV. This selection is so pure that the b-tagging is actually not employed.

The acceptance is typically at the level of 10 to 15\% and
is determined with the simulation (MC), corrected for DATA/MC weight factors.

The backgrounds arise mainly from  
$Z\rightarrow \mu\mu$+jets,
$Z\rightarrow ee     $+jets,
$Z\rightarrow \tau\tau$+jets,
$WW$+jets
These backgrounds are determined with the MC.
Non-$W$ background consists mainly of QCD events in which  jets are faking leptons, it is determined from the data.

The results for the CDF analysis in 750~pb$^{-1}$ of data reads:
  $$\sigma(t\bar t)= 8.3 \pm 1.5 (stat)\pm 1.0 (syst)\pm 0.5 (lumi)\mathrm{\ pb}.$$
For this analysis the main systematic uncertainties arise from the lepton-id (4\%), the JES (3.1\%) and the $t\bar t$ modeling (5\%).

\subsection{Fully hadronic channel}
\begin{figure*}[t]
\centering
\includegraphics[width=75mm]{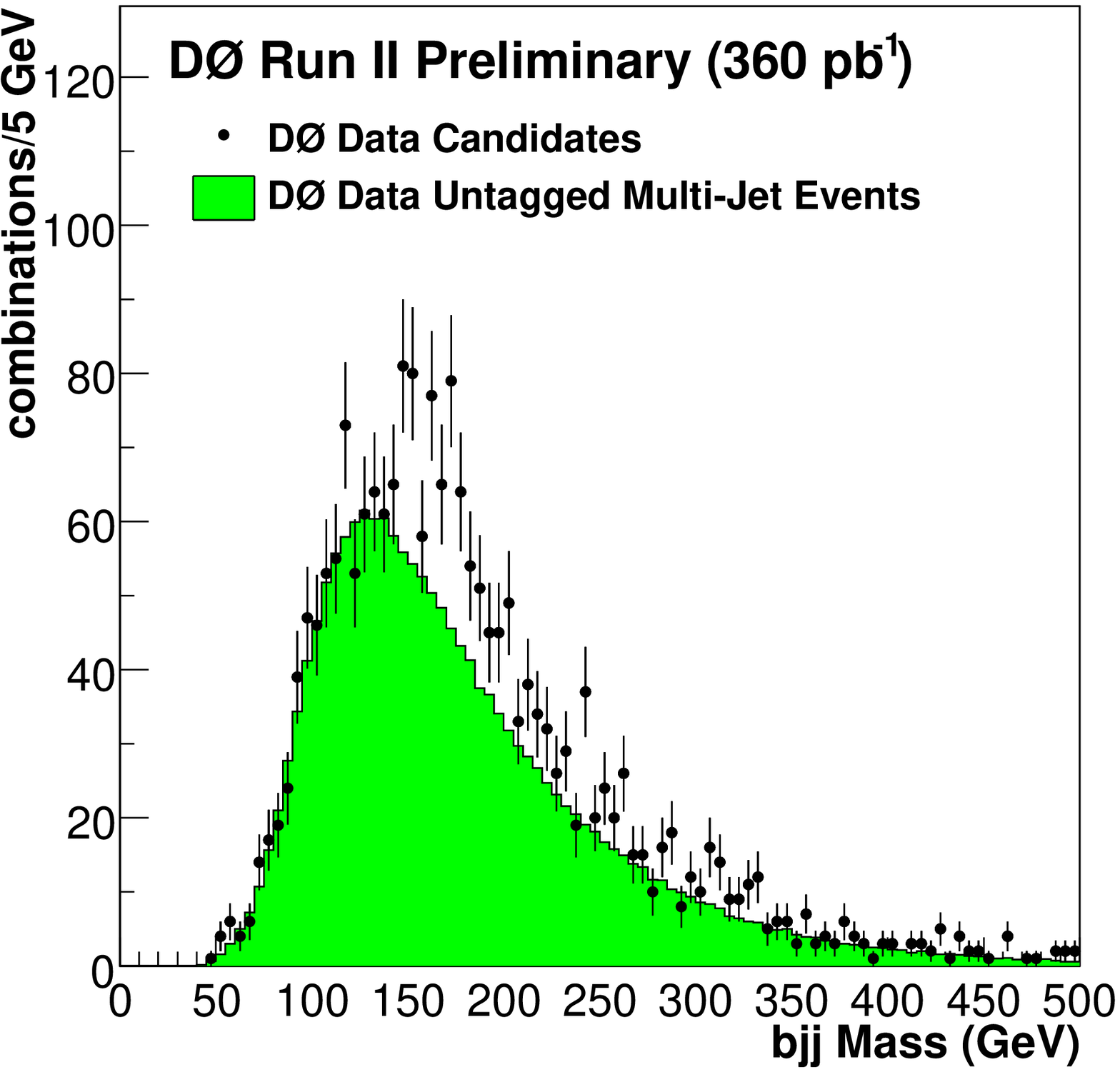}
\includegraphics[width=75mm]{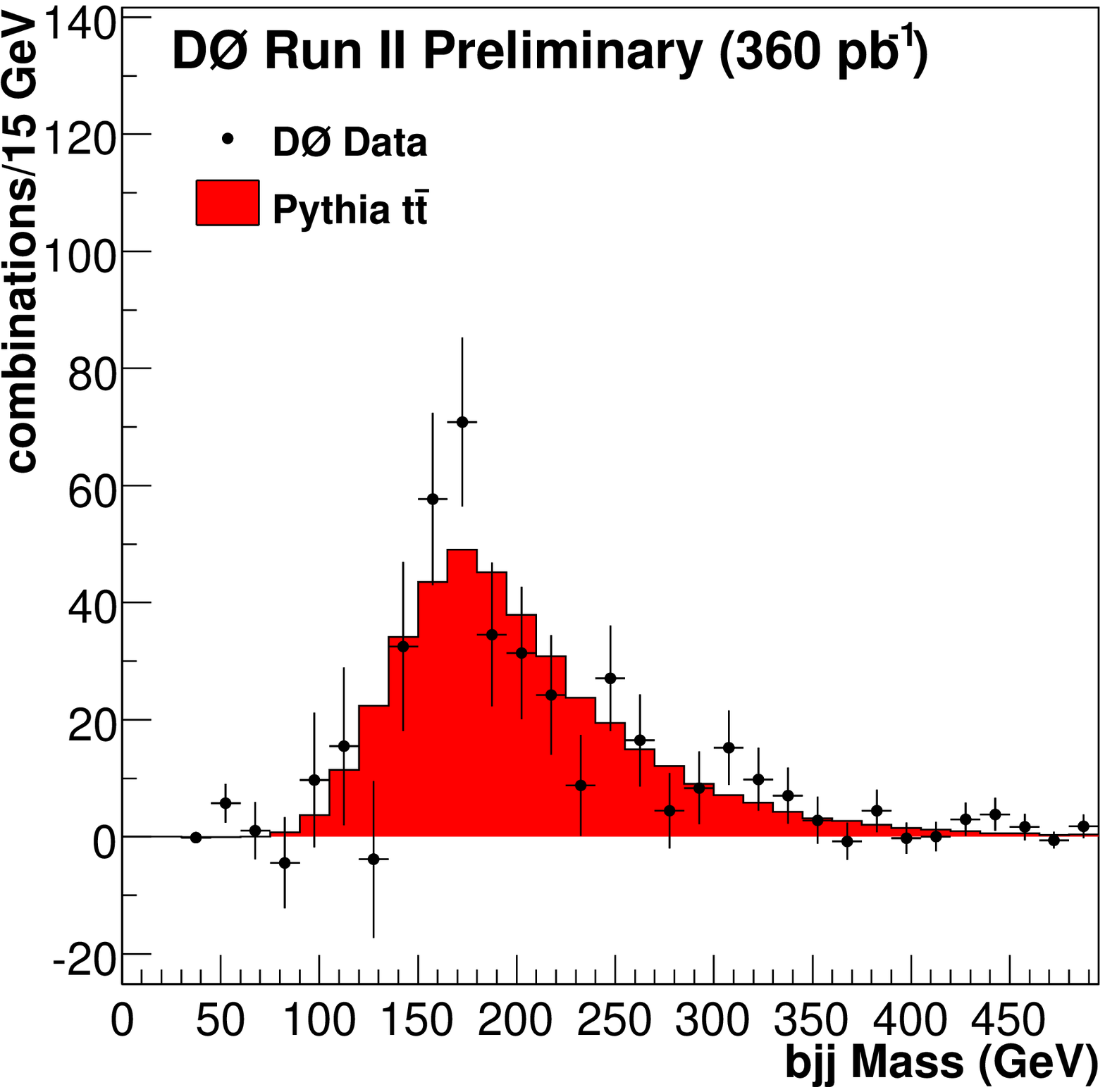}
\caption{Invariant mass of top system ($bjj$), before and after background subtraction in the D\O\ fully hadronic cross-section analysis.} \label{fig:d0_hadronic}
\end{figure*}

The fully hadronic channels suffers from a large QCD background
and a large combinatoric when one tries to match the jets to
the $W$ systems or the $t\rightarrow W b$ system.
The requirement of 2 b-tag jets in the events helps in improving
the signal over background ratio and reducing the combinatorics.
But the background is still very large as shown in Figure~\ref{fig:d0_hadronic}
where the invariant mass spectrum of the reconstructed top system is displayed, before and after background subtraction.
D\O\ with 360~pb$^{-1}$ of data measures
  $$\sigma(t\bar t)= 12.1 \pm 4.9 (stat)\pm 4.6 (syst)\ \mathrm{pb}.$$
The main uncertainties arise from the background subtraction (25\%), the JES (15\%) and the b-tagging efficiency uncertainty (18\%).

\subsection{Pair production summary}

\begin{figure*}[tb]
\centering
\includegraphics[width=75mm]{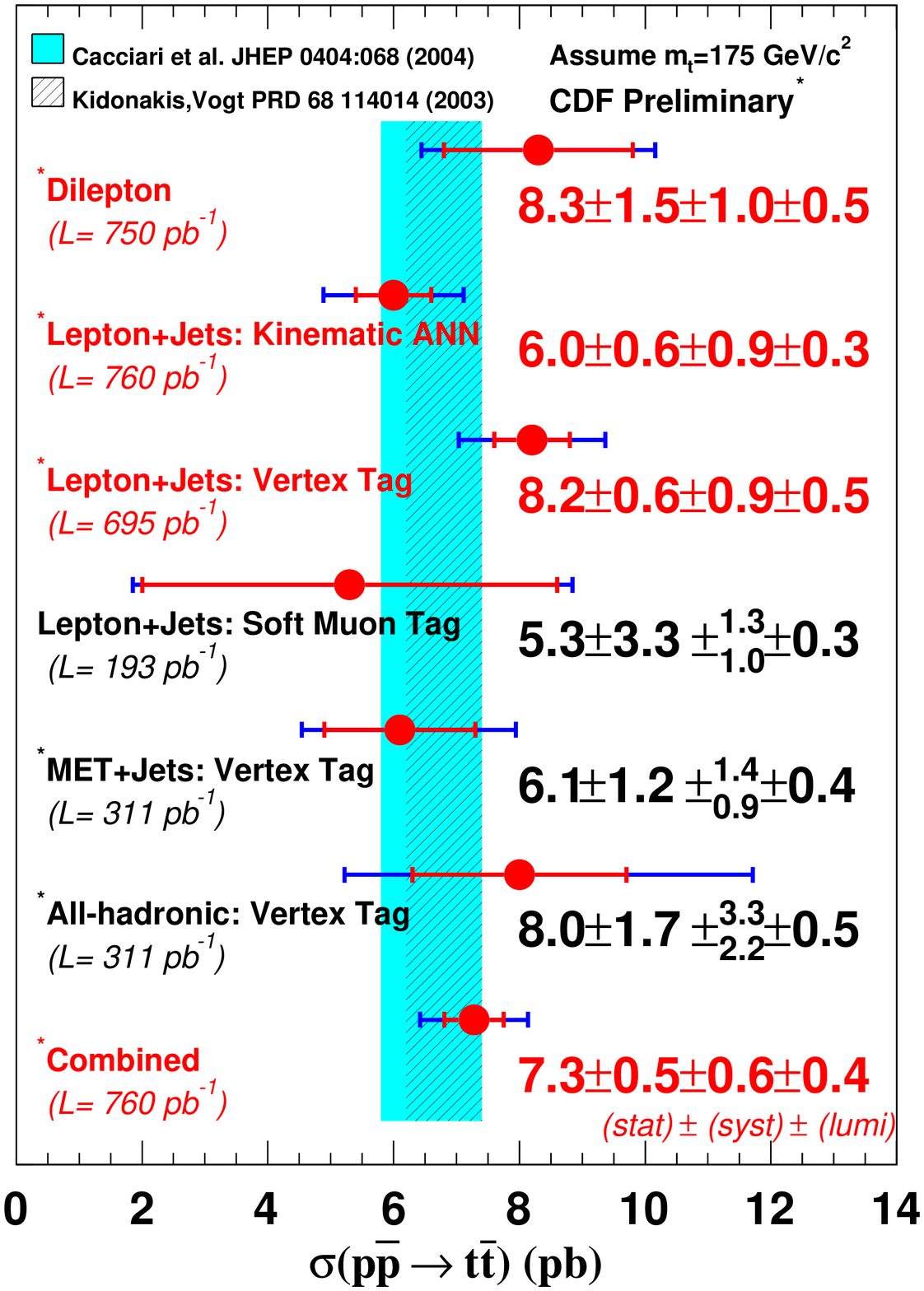}
\includegraphics[width=85mm]{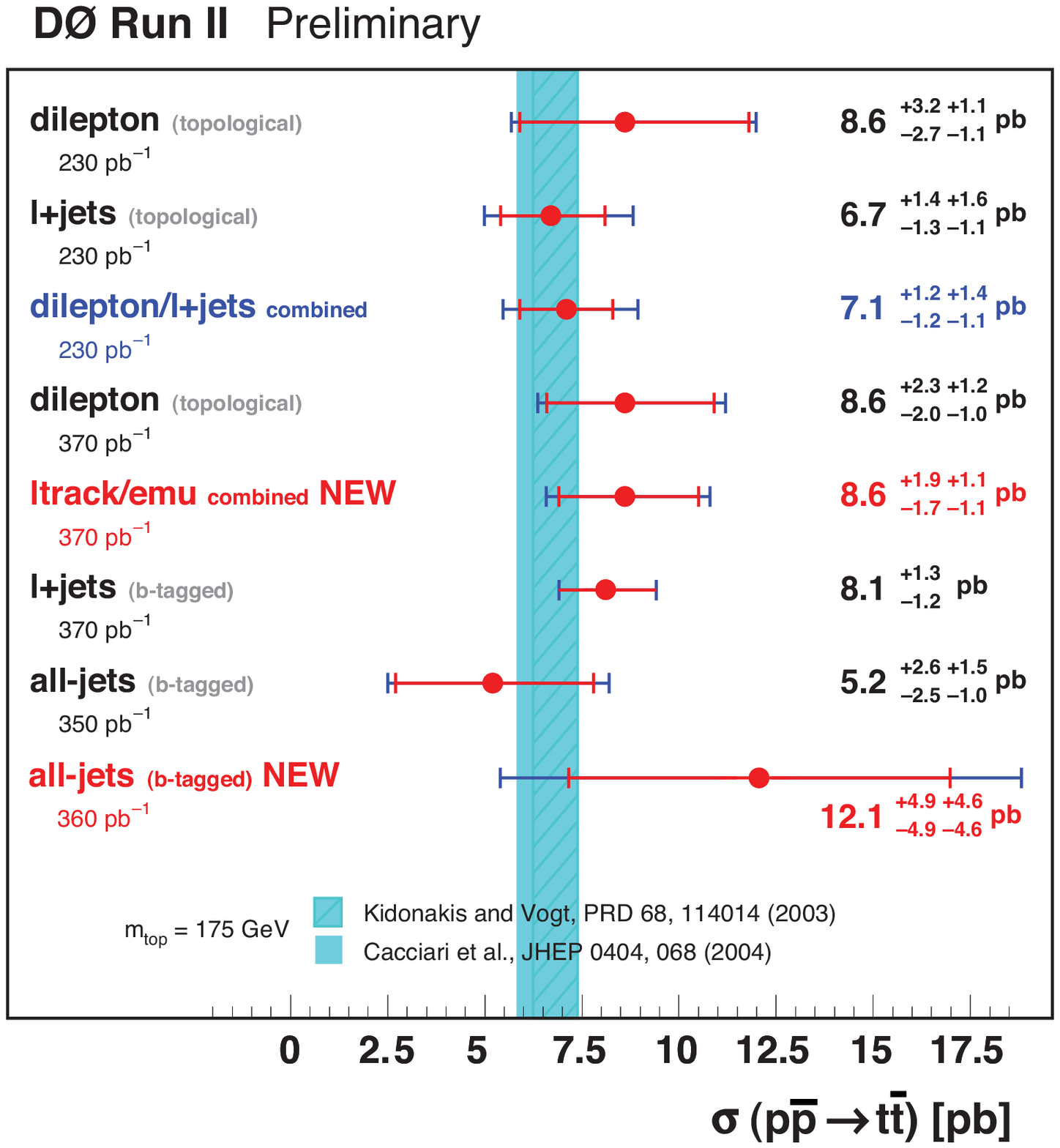}
\caption{cross-sections measured at CDF and D\O\ in the various channels} \label{fig:xsec_summary}
\end{figure*}

The results of the different channels in both CDF and D\O\ are displayed in Figure~\ref{fig:xsec_summary}.
The combined CDF result reads:
 $$\sigma(t\bar t)= 7.3 \pm 0.5 (stat)\pm 0.6 (syst)\pm 0.4 (lumi)\ \mathrm{pb}.$$
It is worthwhile to notice that the level of uncertainty is 12\%, as low as the theoritical uncertainty.
The accuracy is also no more limited by statistics.
Thus it can be said that the Tevatron has entered a precision era regarding the $t \bar t$ cross-section measurement.

\section{Measurement of the top mass}
Measuring the top mass is a complex experimental task.
The events are complicated since
the  $t\bar t$ production yields
six elementary particles in the final state.
Even if the $b-$ jets have been identified, there  is still a large combinatorics when trying to assign reconstructed jets to initial partons.

Besides, these six particles correspond to different type of objects (jets, $b$-tagged jets, leptons, \etmis), which have to be well identified and measured.
As a consequence, the mass measurement depends on the simultaneous understanding of most of the different sub-detectors  (tracking system, electromagnetic calorimetry, hadronic calorimetry, muon system) of CDF or D\O.

Another difficulty comes from the presence of background. Its contamination is sizable so that it   can bias the measurement  if it is not  well understood and under control.

Other sources of difficulty are encountered  when reconstructing the kinematic information in events with \etmis. This is particularly true in the di-leptons channels for which two neutrinos carry out unmeasured energy.

Finally, one of the largest source of experimental difficulty is the parton energy measurement. Resolution effects are typically  $85\%\sqrt E_T$ and go down with increasing statistics. The uncertainty on the jet energy scale is however an irreducible systematics. 
Recently, an  improved method have been set up to reduce this uncertainty:
using an {\em in-situ} calibration of the JES, a part of its
uncertainty scales down with increasing statistics as well. This method will be discussed in the following sections.

\subsection{Methodology}
A lot of different analysis have been performed both at D\O\ and CDF.
They are different from each other depending on:
\begin{itemize}
\item
The channel, lepton+jets, or di-leptons of fully hadronic;
\item the event selection, for example requiring 0 b-tag, 1 b-tag or 2-btag;
\item the jet multiplicity requirements;
\item the method to derive a mass measurement.
\end{itemize}

The various methods can be classified in two  families:
\begin{itemize}
\item
The template methods consist in choosing a given kinematic observable,
creating signal templates at different $M_ {top}$ thanks to the MC,
creating template
for background events, and using a likelihood fit to determine the best signal template;
this method makes use of the kinematic information only.
\item 
The matrix method consists in building a likelihood function based upon the
PDF, the matrix elements of  $t\bar t$ process, and the transfer functions which relate detector measurements (eg: jets \pt) to the  top decay products (eg quarks \pt). Unmeasured quantities are integrated out and a maximum likelihood fit allows to derive the top mass.
This method needs also to be calibrated using MC events.
\end{itemize}

The different analysis do not suffer from the same sources of uncertainty.
This allows to check their consistency  and precision is gained by combining the results.
Only three examples (the two most sensitive and one original analysis) are briefly described in the following.

\subsection {Lepton+jets template method at CDF}
 
The Lepton+jets template analysis at CDF has been performed on 680~pb$^{-1}$ of data. Four exclusive selections with different purities are performed,
namely the 0-btag selection, the 1 b-tag loose selection, the 1 b-tag tight selection and the 2 b-tag selection.
\begin{figure}[t]
\centering
\includegraphics[width=75mm]{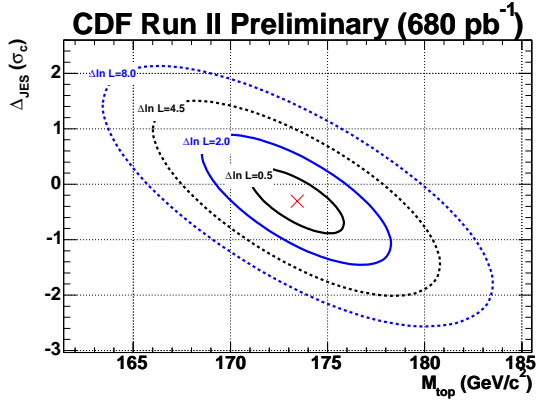}
\caption{Likelihood contour in plane ($M_{top},\Delta JES$) for the CDF template method.}
\label{fig:cdf_mass_contour}
\end{figure}

\begin{table}[t]
\begin{tabular}{|l|c|}
\hline
Systematic Source   &  $\Delta M_{top} (\gevcs)$ \\
\hline
b-jet energy scale   &    0.6 \\
Residual JES        &    0.7 \\
Background JES      &   0.4  \\
ISR                 & 0.5    \\
FSR                 &   0.2       \\
PDF                 &  0.3   \\
Generators          & 0.2 \\
Background Shape    &   0.5 \\
b-tagging           &   0.1   \\
MC statistics       &    0.3  \\
\hline
total  &  1.3 \\
\hline
\end{tabular}
\caption{Systematic uncertainties for the top mass measurement in the lepton+jets channel at CDF.}
\label{tab:cdf_mass_syst}.
\end{table}

In this channel, the six decay products (if eg the $W^+$ decays semi-leptonically) are expected to be:
a lepton $\ell^+$,  a neutrino $\nu$, two $b$-quarks $b \bar b$ and two light quarks $q$ and $q'$.
A kinematic fit allows to choose among the various combinations, making use of the b-tag, which reconstructed jets correspond the the $qq'$ system, which jets
correspond to the hadronic top, $qq'\bar b$, and which jet should be associated to the ``leptonic top'', $b\ell^+\nu$.
Thus for each event a top mass, $M_{qq'\bar b}=M_{l\nu b}$, and   a $W$ mass, $M_{qq'}$, are  reconstructed.

Templates for these two quantities are prepared from the MC after varying the
top mass hypothesis, $M_{top}$, and $\Delta JES$ which is defined as the deviation from the usual JES calibration in unit of uncertainty on this calibration.

By comparing the templates to the data, a simultaneous fit of ($M_{top},\Delta JES) $ is performed.
Results of this fit are displayed in Figure~\ref{fig:cdf_mass_contour}.
The results of this fit are the world best single measurement, at the time of PIC06. They read
$$ M_{top}=173.4 \pm 2. 5(stat+\Delta JES) \pm 1.3 (syst) \ \gevcs.$$

The measured $\Delta JES$ is  $$\Delta JES=-0.3\pm 0.6 (stat+M_{top}).$$
This signifies that the systematic uncertainty due to the JES as been scaled by 0.6 thanks to the in-situ calibration and the $W$ mass constraint. A further decrease is expected in the future with more statistics.

The systematics on the mass measurement are summarized in Table~\ref{tab:cdf_mass_syst}.

\subsection {Lepton+jets  matrix element method at D\O\ }

The D\O\ collaboration has analyzed 380~pb$^{-1}$ of data using
a matrix element method in the lepton+jets channels.
Three exclusive samples are selected, namely a 0 b-tag, 1 b-tag and 2 b-tag samples.
The signal probability for a set of measurements (jets, leptons) $\vec x$ is built:
\begin{eqnarray}\label{eq:ME}
P(\vec x, M_t, JES )=& \\ 
\frac 1 \sigma \int d\sigma(\vec y, M_t)  & f_1(\tilde q_1)f_2(\tilde q_2)
&W(\vec x, \vec y,  JES)d\tilde q_1d \tilde q_2 , \nonumber
\end{eqnarray}
where $\tilde q_i$ are the initial parton energies, $f_i(\tilde q_i)$ are the proton (and anti-proton) parton density function,
$W$ is the transfer function from true kinematic quantities $\vec y$ to measured $\vec x$  and $d\sigma$ is the differential cross-section.

In Equation~\ref{eq:ME}, $JES$ is an additional scaling factor to the externally calibrated Jet Energy Scale. It is left as a free parameter, so that
it can be simultaneously fitted, together with the top mass $M_t$. Thus an in-situ calibration is performed, mostly relying on the constraint arising from  the $W\rightarrow q\bar q'$ system.

In a similar way, a probability is built for background events $P_{bkg}$ and the final likelihood reads: 
\begin{equation}\label{eq:Likelihood_ME}
\mathcal L = \prod_{evt}( f_{top} P(\vec x, M_t, JES )+ (1 - f_{top} )P_{bkg}(\vec x, JES))
\end{equation}

\begin{figure}[t]
\centering
\includegraphics[width=75mm]{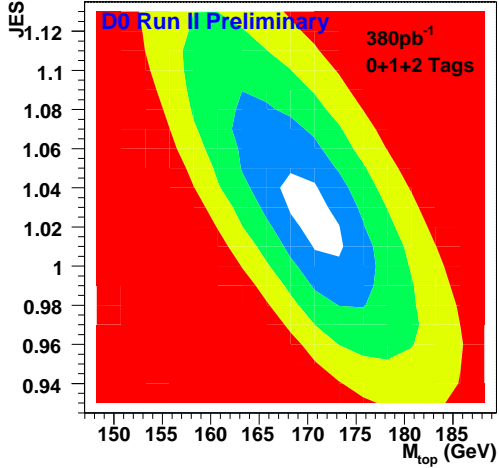}
\caption{Likelihood contour in plane ($M_{top}, JES$) for the D\O\ matrix element method. The different colored region correspond to steps $\Delta \mathcal L=0.5$.}
\label{fig:d0_mass_contour}\end{figure}

The Likelihood function is shown in Figure~\ref{fig:d0_mass_contour}.

The results of the fit reads:
$$
 M_{top}=170.6^{+0.4}_{-4.7}(stat+\Delta JES) \pm 1.4 (syst) \ \gevcs.
$$
The JES in-situ calibration gives:
$$
JES =1.027^{+0.033}_{-0.030}(stat+M_{top}) .
$$
it indicates a 2.7\% correction relative to the externaly calibrated JES, well compatible with its uncertainty ($\simeq\mathrm{ 3\ to\ 4}\%$). The error on the fit reads $^{+0.033}_{-0.030}$. When compared to the externally calibrated JES uncertainty this indicates a small improvement thanks to the in-situ calibration. But this improvements will be more significant with a higher statistics.

The systematics on the mass measurement are summarized in Table~\ref{tab:d0_mass_syst}.

\begin{table}[t]
\begin{tabular}{|l|c|}
\hline
Systematic Source   &  $\Delta M_{top} (\gevcs)$ \\
\hline
b-fragmentation    &    0.7 \\
b-response          &    0.8 \\
MC calibration       &   0.45 \\
$W$ background      & 0.3    \\
QCD background     &   0.3       \\
total  &  1.4 \\
\hline
\end{tabular}
\caption{Systematic uncertainties of the top mass measurement at D\O\ in the lepton+jets channel with the matrix element method.}
\label{tab:d0_mass_syst}.
\end{table}

\subsection{Decay length method}
This method employed at CDF is very different from the other ones,
as it does not rely on the energy measurements of outgoing particles.
Thus it is an important cross-check with reduced impact from the JES uncertainty.

The method  makes use of the B hadron decay length and the 
boost of the b-quark in top decays, directly related to the top mass:
\begin{equation}
\gamma_b =\frac 
{M_t^2+M_b^2-M_W^2}
{2M_tM_b} \simeq 0.4 \frac {M_t}{M_b}.
\end{equation}

In a lepton+jets sample selected from 695 \invpb\ of data, the decay length of the b-hadrons are reconstructed and compared to template simulated with different top mass hypothesis.
The result reads:
$$
 M_{top}=183.6^{+15.7}_{-13.9}(stat)\pm 0.3 (JES) \pm 5.6 (syst) \ \gevcs.
$$

As expected, the impact of JES uncertainty is very small,
but the result is statistically limited using the present data sample. However such kind of analysis will become competitive at the end of Run II, when ten times more data will be recorded.


\subsection{Combined results}
Nine measurements performed at CDF and D\O\ have been combined to achieve
a 1.3\% accuracy:
$$
 M_{top}=172.5\pm 1.3 (stat)\pm 1.9 (syst)\ \gevcs.
$$

\begin{figure}[t]
\centering
\includegraphics[width=0.45\textwidth]{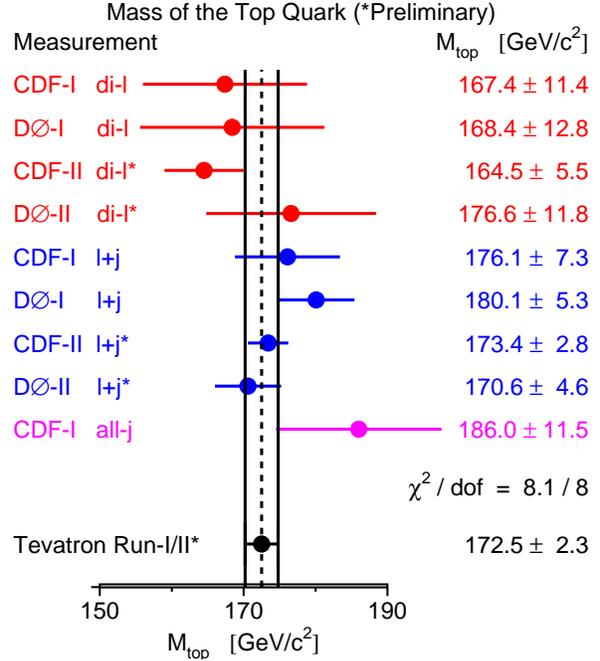}
\caption{Measurements of the top mass at CDF and D\O.}
\label{fig:cdf_d0_mass_summary}
\end{figure}

The various measurements are summarized in Figure~\ref{fig:cdf_d0_mass_summary}.

The chi square of the combination is $\chi^2/dof = 8.1/8$ which demonstrate a good consistency.

It should be stressed that the Run~IIa goal was to obtain a 2.5~\gevcs\ combined uncertainty using $\simeq 2\ \invfb$ of data. This goal has been exceeded. The 1\% precision is in sight for the near future. 

\subsection{Impact on the Higgs Boson Mass}

 \begin{figure}[t]
\centering
\vspace{-3cm}
\includegraphics[width=0.45\textwidth]{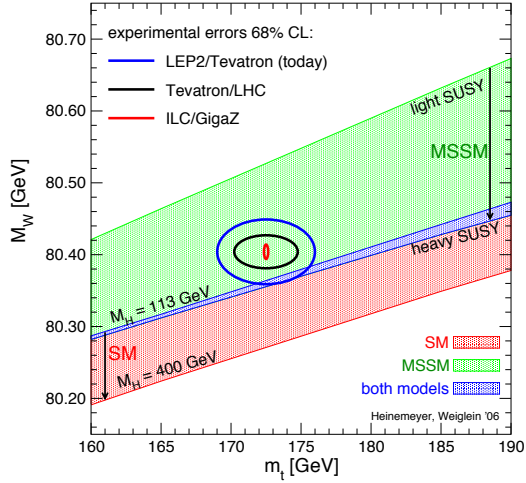}
\caption{
Error ellipse projections for the top and $W$ masses from current LEP2 and Tevatron data and for future Tevatron/LHC  and ILC sensitivities. These are overlaid onto bands of predicted Higgs masses for the SM, as well as for different SUSY scenario~\cite{susy_oversm}.
}
\label{fig:mt_vs_mw}.
\end{figure}
The top mass is a  free parameter in the standard model.
But its mass can be related to various electroweak observables thanks to radiative corrections. Thus the  measurement of the top mass provides a consistency check of the SM that can be used to probe for new physics.

In particular the top mass plays an important role in the radiative corrections to the $W$ boson propagator: $\Delta M_W \alpha M_t^2 $. As the Higgs boson mass also influences the $W$ boson mass through radiative corrections, $\Delta M_W \alpha \log(M_H) $, the combined measurements of the $W$ and top masses yield constraints onto the SM  Higgs boson mass.

Using the top mass combined  measurement presented in the previous section, the SM Higgs boson is predicted to have a mass fulfilling (spring 2006):
$$
 M_H= 89 ^{+42}_{-30}\ \gevcs ,
$$
and the upper bound at 95\% Confidence Level (C.L.) is:
$$
 M_H< 175\ \gevcs\  
$$
  It is worthwhile to notice that our present data tend to favor Supersymmetry over the Standard Model~\cite{susy_oversm} as shown in  Figure~\ref{fig:mt_vs_mw}.

\section{Top quark properties}
Other properties than the top mass can be measured at Tevatron to check the SM consistency and search for new physics.

\subsection{Top charge}
Within the SM, the top quark charge is expected to be $+\frac 2 3$, but one can imagine model in which a heavy quark exists  that carries a different charge.
For example as in \cite{Chang:1998pt} one can imagine an exotic right doublet $(Q_1,Q_4)_R$ of charge $(-\frac{1}3,-\frac 4  3)$, $Q_4$ being the heavy quark observed at Tevatron.
In the semi-leptonic mode, the top quark of charge $+\frac 2 3$ decays into $\ell^ + \nu \bar b$ while a quark of charge $+\frac 4 3 $ would decay into  $\ell ^+ \nu b$ .

The D\O\ collaboration has analyzed 370~\invpb\ of data to select a pure lepton+jets+2 b-tag  sample.
To associate to right b-jet to the (lepton,\etmis) system a fit to the top mass is performed.

The charge of a b-jet is defined by weighting the contribution of the charged tracks matched to the jets:
\begin{equation}
Q_{jet}=\frac 
{\sum_i q_i.p_{Ti}^\kappa}  
{\sum_ip_{Ti}^\kappa}  
 \ \mathrm{,\ with\ \kappa=0.6.}
\end{equation}

\begin{figure}[t]
\centering
\includegraphics[width=0.5\textwidth]{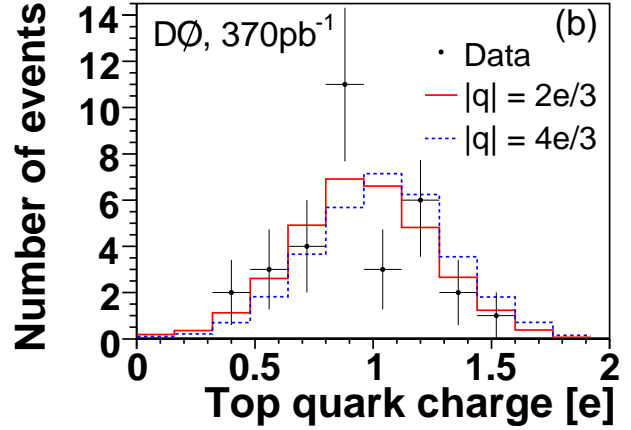}
\caption{Reconstructed top charge in lepton+jets events at D\O.}
\label{fig:d0_top_charge}
\end{figure}

Two observables allow to measure to top charge twice per event:
$\tilde Q_1=|q_b+q_\ell|$ and
$\tilde Q_2=|-q_B+q_\ell|$, where $b$ is the b-jet matched to the lepton and $B$ is the other b-jet.
A fit of MC templates to the data distribution of $\tilde Q_1$ and $\tilde Q_2$ is performed, which yields the exclusion of $Q=\frac 4 3$ at the 93.7\% confidence level. The distribution is shown in Figure~\ref{fig:d0_top_charge}.

\subsection{Top lifetime}
The measured top lifetime could be significantly different from the SM value, $c\tau=3.10^{-10} \ \mu\mathrm m$, if for example a long lived decaying particles is involved in the top production. A top lifetime analysis has been performed at CDF using 318~\invpb\ of data in a lepton+jets+1~b-tag selection. The impact parameter of the lepton is used as an observable related to the top decay length. A maximum likelihood fit of  the impact parameter from data to  MC templates yields the exclusion of long decay length:
$c\tau < 52.5 \ \mu\mathrm m$ at 95\% confidence level.

\section{Wtb vertex}
Studying the $Wtb$ vertex is an important test of the SM which probes the 
$(V-A)$ nature of the weak interaction and also checks the unitarity of the CKM matrix.

\subsection{Measurement of Vtb}
Within the SM the CKM matrix element is involved in the top branching fractions:
\begin{equation}
R= \frac
{BR(t\rightarrow Wb)}
{BR(t\rightarrow Wq)}
=\frac
{|V_{tb}|^2}
{|V_{tb}|^2+|V_{ts}|^2+|V_{td}|^2} =
\begin{array}{r}
0.9980\ \\\mathrm{to\ } 0.9984
\end{array}
\end{equation}

\begin{figure}[t]
\centering
\includegraphics[width=0.45\textwidth]{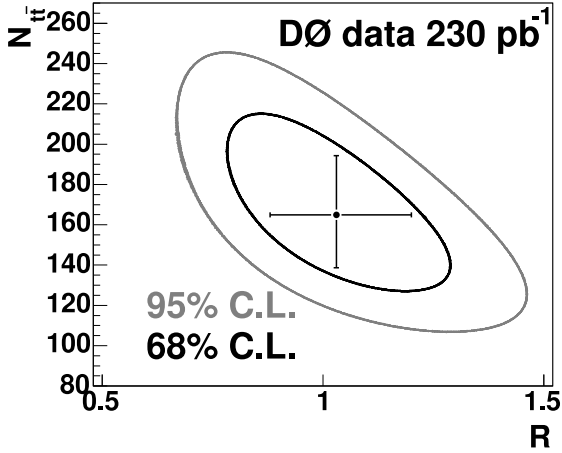}
\caption{Simultaneous measurement of top cross-section and branching fraction to $Wb$ in the D\O\ experiment.}
\label{fig:d0_top_br}
\end{figure}

To measure this branching fraction, D\O\ has selected a sample of lepton+jets events in 230~\invpb\ of data.
By counting the number of selected events and the number of b-tag jets,  a simultaneous  determination of $R$ and the $t\bar t$ cross-section is performed.
Since there is no a priori regarding the  b-jet content of the events, a model indepent cross-section is obtained. The results of the 2d fit are displayed in Figure~\ref{fig:d0_top_br}. They are well in agreement with the Standard Model expectations:
$$
\sigma(t \bar t)=7.9^{+1.7}_{-1.5}\mathrm{\ pb}
$$
$$
\frac
{BR(t\rightarrow Wb)}
{BR(t\rightarrow Wq)}
=1.03^{+0.19}_{-0.17}
$$

The same kind of studies has also been conducted at CDF.
Using 162~\invpb\ the CDF collaboration also measures a branching fraction in agreement with the SM:
\begin{equation}
\frac
{BR(t\rightarrow Wb)}
{BR(t\rightarrow Wq)}
=1.12^{+0.21}_{-0.19} (stat)^{+0.17}_{-0.13} (syst)
\end{equation}

\subsection{Measurement of W helicity}
In the Standard Model, the $(V-A)$ nature of the electroweak interaction is also present in the $Wtb$ vertex. As a consequence,  almost $F^+\simeq 0\%$ of the $W$ arising from the top decays should have a right-handed helicity.
One also expects   $F^
0\simeq 70\%$ of longitudinal $W$ and  $F^-\simeq 30\%$ of left-handed $W$.

By studying the $W$ decay product kinematics, it is possible to measure the $W$ helicity and probe $(V-A)$ structure of the electroweak interaction at energy close to the electroweak symmetry breaking scale.
Several  variables can be used to relate the event topology to the $W$ helicity: the \pt\ of the leptons, the decay angle of the leptons in the top rest frame, $\cos\theta^\star$, or the  invariant mass  $M^2_{lb}=\frac 1 2 (M_t^2-M_W^2)(1+\cos\theta^\star)$.

A fit to the $M^2_{lb}$ spectrum observed from  lepton+jets and di-lepton top events with 695\invpb\ at CDF yields a result perfectly compatible with the SM:
$$
F^+= 0.02\pm0.07
$$
$$F^+< 0.09 \mathrm{\ at\ 95\%\ C.L.}
$$

At D\O, using 370~\invpb\ of data, the study of the $\cos\theta^*$ spectrum in lepton+jets and di-lepton samples also shows a good agreement with the SM:
$$
F^+= 0.08\pm0.08 (stat)\pm 0.06 (syst)
$$
\begin{figure*}[t]
\centering
\includegraphics[width=0.48\textwidth]{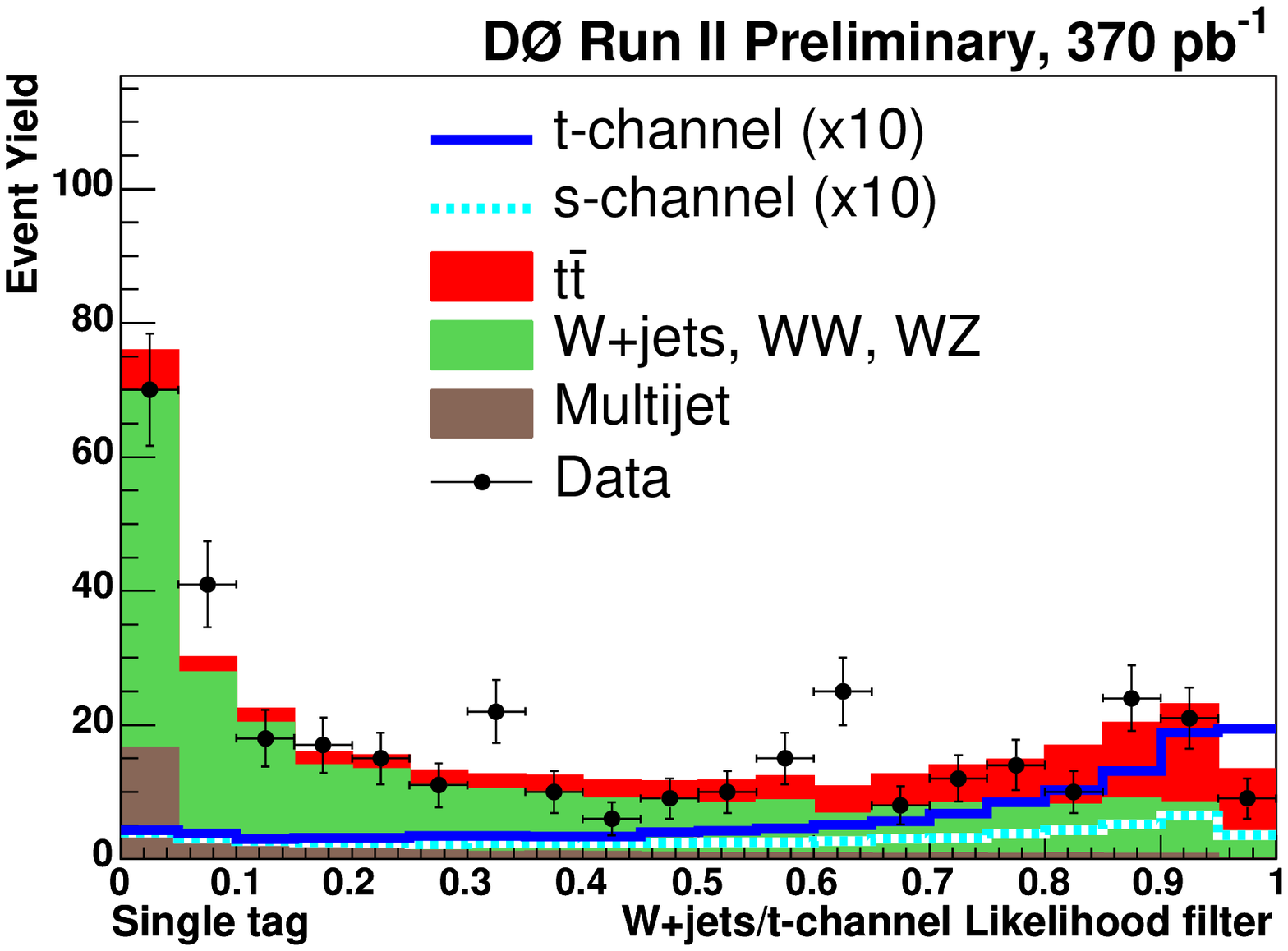}
\includegraphics[width=0.48\textwidth]{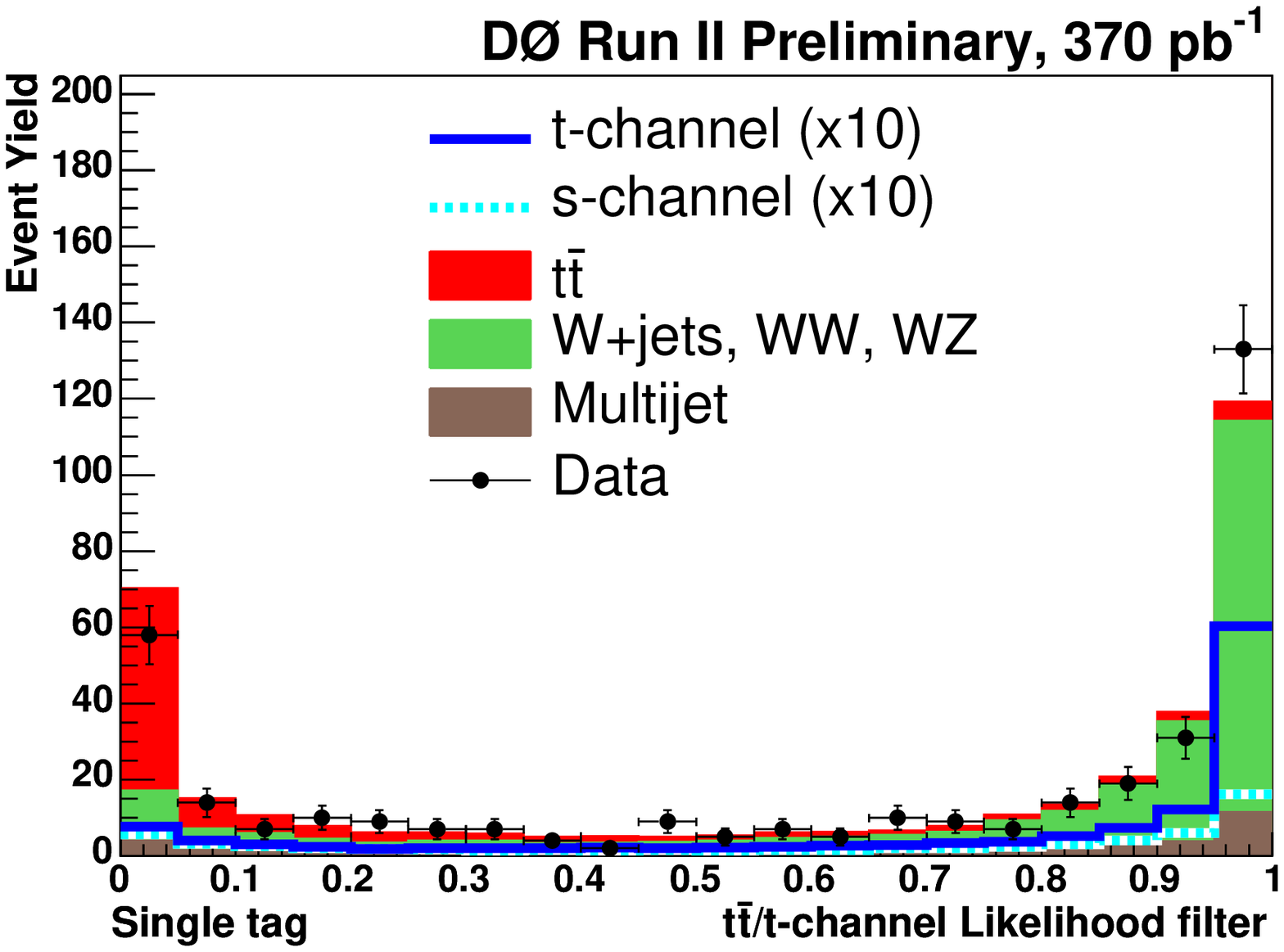}
\caption
{t-channel discriminant distributions for single top searches at D\O: on the left plot: $W+jets$ filter, on the right plot: $t\bar t$ filter. 
}\label{fig:d0_singletop_disc}
\end{figure*}
\subsection{Search for single top}
As mentioned in Section~\ref{sec:singletop}, the production of single top involves the $Wtb$ coupling in two different channels:
the s-channel gives rise to a $\ell\nu b\bar b$ final state with kinematic properties different  from the t-channel and its $\ell\nu q b$ final state.
The high background arises from W+jets, especially W+$b$-jets, and also from the $t\bar t$ production. To discriminate against the background, multivariate techniques are used both at D\O\ and CDF: ANN, Likelihood or decision trees.
For example at D\O, 4 likelihood variables are built to discriminate:
t-channel against $t\bar t$, t-channel against $W+jets$, s-channel against $t\bar t$ and s-channel against $W+jets$.
Figure~\ref{fig:d0_singletop_disc} shows two out of the four possible distributions for these likelihood variables.
Given the present luminosity, the single top has not been observed in the data and both D\O\ and CDF have set limits on the cross-section, which are in agreement with the SM expectations:
at D\O\, the results arising from the likelihood analysis with 370~\invpb\ of data are:
\begin{equation}
s:\ \sigma<5.0\ \mathrm{pb\ at\ 95\%\ C.L.}
\end{equation}
\begin{equation}
t:\ \sigma<4.4\ \mathrm{pb\ at\ 95\%\  C.L.}
\end{equation}
At CDF with 695~\invpb\ of data, an ANN analysis yields the results:
\begin{equation}
s:\ \sigma<3.2\ \mathrm{pb\ at\ 95\%\ C.L.}
\end{equation}
\begin{equation}
t:\ \sigma<3.1\ \mathrm{pb\ at\ 95\%\  C.L.}
\end{equation}
\begin{equation}
s+t:\ \sigma<3.4\ \mathrm{pb\ at\ 95\%\ C.L.}
\end{equation}

Prospect studies have been performed at CDF, which indicate that with 1.5~\invfb, the single top could be observed at the $3\sigma$ level.
On one hand these studies do not account for the systematic uncertainties which could delay the discovery by a few months. On the other hand, they do not account for the possible improvements in analysis techniques. Including a safety margin, we conclude that the  evidence of single top production should be observed by the end of 2006 or 2007 at both CDF and D\O.

\section{Search for new particles}
\subsection{Heavy top: t'}
If a $\mathrm{4^{th}}$ generation heavy quark exists, it may show up just like the top quark, but with different kinematic properties. Namely the events will be produced with higher reconstructed mass, and $H_T$. CDF has searched
for a heavy particle in the lepton+jets sample, selected from  760~\invpb\ of data, without the use of  b-tag requirement. The discriminant quantities, $M_Q$ and $H_T$ are reconstructed to probe for a possible heavy quark production. No excess is observed in the distribution of these variables, so that a limit is set on the $t'$ mass.
$$
M_{t'} > 258\ \gevcs\ \mathrm{at }\ 95\%\ \mathrm{C.L.} 
$$
 
\subsection{Resonance in top production}
New physics may shows up as a resonance in the production of top pair.
The events are selected in a lepton+jets sample and the invariant mass
of the $t\bar t$ pair is reconstructed.
No excess is observed by either D\O\ in 370~\invpb\ or CDF in 680~\invpb, so that limit can be put on the resonant production.
In a Lepto-phobic top-color model, the heavy $Z'$ is excluded following
$$
M_Z' > 725\ \gevcs\  \mathrm{at }\ 95\%\ \mathrm{C.L.\  (CDF\ 680~\invpb)}
$$
$$
M_Z' > 680\ \gevcs\  \mathrm{at }\ 95\%\ \mathrm{C.L.\  (D0 \ 370~\invpb)}
$$

\subsection{Search for W'}
A leptophobic $W$-like boson would show up as a resonance producing single top.
To search for this kind of particle, D\O\ employs the same techniques as for the SM single top: in the lepton+jets channel an ANN is built to discriminate against both the $t\bar t$ background and the the $W$+jets background.
The production of single top would be signed as a peak in the distribution of the reconstructed invariant mass, $\sqrt{\hat s}$, of the selected events.
As no excess is observed in the 230~\invpb\ of analyzed data, limits are set on the mass of the $ W'$.
$$
200<M_W' < 650\ \gevcs\  \mathrm{ excluded\ \mathrm{at }\ 95\%\ {C.L.\  }
}
$$
\subsection{Search for charged Higgs}
\begin{figure}[t]
\centering
\includegraphics[ width=0.48\textwidth]{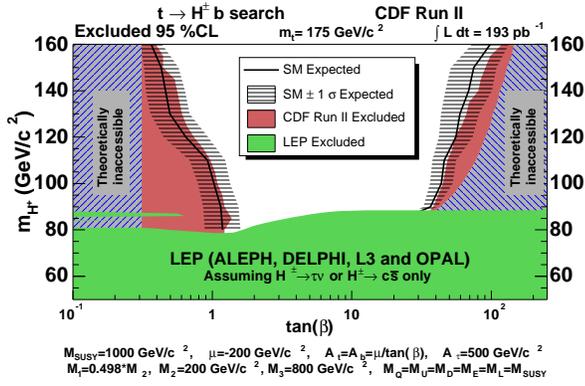}
\caption
{Example of benchmark scenario for which exclusion in the plane $(M_{H^+}, \tan\beta )$ are derived}
\label{fig:cdf_mssm_hplus_exclusion}
\end{figure}

\begin{figure}[t]
\includegraphics[ width=0.48\textwidth]{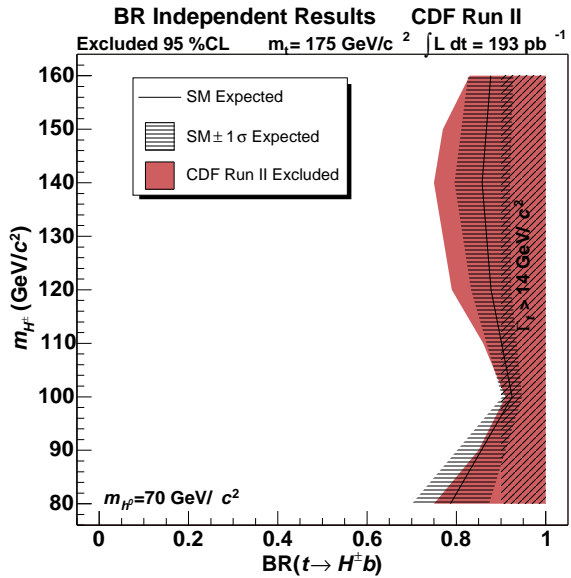}
\caption
{Model independent exclusion limit in the plane ($m_H^+,Br(t\rightarrow H^+b))$ 
}\label{fig:cdf_mssm_hplus_mi_exclusion}
\end{figure}
In two Higgs doublet models, five physical Higgs bosons are predicted: 2 CP-even neutral, $h$ and $H$, one CP-odd neutral, $A$, and two charged, $H^+$,$ H^-$.
The coupling of the charged Higgs to fermions is a function of $\tan\beta$, the ratio of the vacuum expected values of the two doublets. At large or low $\tan\beta$ the coupling to the top is large, so that the decay $t\rightarrow  H^+ b$ is allowed if $H^+$
is light enough.
The possible decays of the charge Higgs boson depend upon, $M_{H^+}$, $\tan\beta$ but also on other parameters such as $M_A$ (decay $H^+\rightarrow AW^+$) or $M_h$  ($H^+\rightarrow hW^+$). 
To search for charged Higgs it  is preferred to use a more constrained model, such as the Minimal Supersymmetric Standard Model (MSSM) for which 2 parameters only describe the Higgs sector at tree level, for example  ($M_{H^+}$, $\tan\beta$). 
Several benchmark scenarios are also well defined to probe the effects of the radiative corrections.

CDF has conducted the search for charged Higgs bosons by looking at the final state of $t\bar t $ production.
For a given set of MSSM parameters, the branching fractions such as
$Br(H^+\rightarrow \tau \nu),$ $ 
Br(H^+\rightarrow  c\bar s),$ or $
Br(H^+\rightarrow  hW^+)$ are well defined. Each point in the MSSM parameter space predicts a given excess or deficit with respect to the SM expectations for the lepton+jets, di-leptons,  single or double b-tag or  $\tau $+jets  channels. 
As no departure from the SM expectation has been observed, limits have been set on the MSSM parameter space.
An example of such a limit is shown in Figure~\ref{fig:cdf_mssm_hplus_exclusion}.

A model independent limit can also be obtained as a function of only $m_H^+$ and the branching fraction
$Br(t\rightarrow H^+b)$. It is displayed in Figure~\ref{fig:cdf_mssm_hplus_exclusion}.
\vfill

\section{Conclusion}
Thanks to the well performing Tevatron, thousands of top quarks have been produced at  both CDF and D\O. The top quark physics program  of Run II is well underway.

The Tevatron experiments have demonstrated their ability to measure various properties of the top quark and probe the Standard Model.  Regarding the mass and cross-section measurements it is clear that we have entered a precision era in which statistical uncertainty is no more dominating.
Regarding physics beyond the Standard Model, top-like final states provide
the opportunity to probe energies close to the electroweak symmetry breaking scale.

By the end of Run IIb in 2009, ten times more data should have been recorded. A bunch of exciting new results, some of them might be unexpected, will have  been released just before entering the LHC era, in which top quarks will be produced in significantly larger quantities.

\bigskip 

\end{document}